# ANALYSIS OF WOVEN REINFORCEMENT PREFORMING USING AN EXPERIMENTAL APPROACH

G. Hivet [1], S. Allaoui [1], D. Soulat [1], A. Wendling [1], S. Chatel[2]

[1] Institut PRISME/MMH, Université d'Orléans, 8 rue Léonard de Vinci 45072 Orléans Cedex 2, France

[2] EADS France, Innovation Works, Mechanical Modelling Research Team, 12 rue Pasteur, BP76, 92152 Suresnes Cedex, France

Gilles.Hivet@univ-orleans.fr ; Damien.Soulat@univ-orleans.fr;
Samir.Allaoui@univ-orleans.fr ; sylvain.chatel@eads.netuthors;
audrey.wendling@estaca.eu

## SUMMARY

In order to understand the mechanisms involved in the forming step of LCM processes and provide validation date to numerical models, a specific experimental device has been designed in collaboration between PRISME Institute and EADS. The goal of this paper is to present the potentialities and the first results obtained with this device.

## INTRODUCTION

Among the manufacturing processes, Liquid Composites Moulding are undoubtedly among the most interesting for the obtaining of a composite part. It offers several advantages like reduced solvent emissions, part quality, and process repeatability [1]. Furthermore, forming processes have shown their efficiency for the manufacturing of metallic materials for many years, so that they have replaced, in many cases, historical processes. Nevertheless, its development in industrial application is not so common today especially because the process has to be fairly improved.

LCM processes start with the forming of a dry fabric on what can be a complex (double curved) shape. This first step is not completely mastered today what constitutes a main drawback for the development of this process at the industrial level. Nowadays, the main industrial approach used to study part feasibility is trial and error. Other approaches have to be considered in the future. Among the strategies useable to investigate a priori the formability of a given fabric on a given shape, finite element simulation [2] and experimental demonstrators can be considered. The complementarity between these two methods will enable to understand and model accurately the performing step and will hopefully permit to decrease the cost and time needed for the tools and fabrics development.

For aeronautical applications, the use of the RTM process is under study in order to obtain a three-dimensional double curved shape for thick parts constituted by several plies of fabric and without defects. Obviously, defects on the dry fabric preform will lead to a significant decrease of the composite performance what is completely unacceptable for flying applications. This paper deals with the design of a specific demonstrator dedicated to the study of this industrial case; it has been developed with the collaboration of EADS.

After a short presentation of the device principles, the expected fabric behaviour is analysed and the mechanical properties of the fabric are measured. Then, the first results concerning the formability of one ply of an interlock fabric on a square part and a tetrahedron are presented.

**DEVICE PRINCIPLES**

**Intents**

The mechanical behaviour of fibres reinforced composites is piloted by the orientation and position of the fibres along their direction, and, along the others, by the homogeneity of the resin. For a fabric (woven or non crimp), three types of results are more specifically interesting:
- the membranous orientation and position of the yarns of one layer with respect to the final shape because this will give the orientation of the stiffest direction of the obtained anisotropic layer,
- the presence of defects of any kind concerning the fabric (fiber or yarn breakage, wrinkles …), because they will reduce significantly the quality of the final part.
- the presence of any defects in the polymerized resin (voids…) for the same reasons.

Understanding the phenomenon that drive the placement of the yarns of a given fabric on a given shape is a crucial point in order to improve the design process for tools and fabrics.

Even if the injection phase might modify the position of the fibres, the principle of the LCM processes implies that the fibres position and orientation are mainly imposed by the first forming step of the dry fabric. Furthermore, it has been demonstrated in many studies [3-5] that the modification of the fibre orientation and local variations of fibre volume fraction during the forming process, have a significant impact on the resin injection step and thus on the defects.

For that purpose, a forming demonstrator should then enable to give:
- the membranous position and orientation of the yarns
- the presence of defect of any type
- the 3D strains of the fabric
- the process parameters

, in function of the process parameters: blank holder force, punch velocity, shape of the tools, fabric, number and orientation of the layers at the initial state.

On these conditions, the demonstrator will allow us to understand and model the link between the mechanical properties of the fabric, the process parameters and the final placement of the fibres. The experimental results obtained will also be useful to improve the quality of numerical prediction [2]

**Constitution of the device:**

The demonstrator is composed of three parts. The first part consists in a pair punch/open die, the die is chosen open in order to see perfectly the yarns during the stamping. The punch is moved using an electric jack in order to control easily and accurately the punch position and speed. The device is build so that any pair of punch/die can be adapted. Since our study concerns a corner of a square box, at the moment three couples have been manufactured: a square box, a tetrahedron, and a prism (figure 1.) Sensors enable to track: the position, speed and load exerted by the punch.

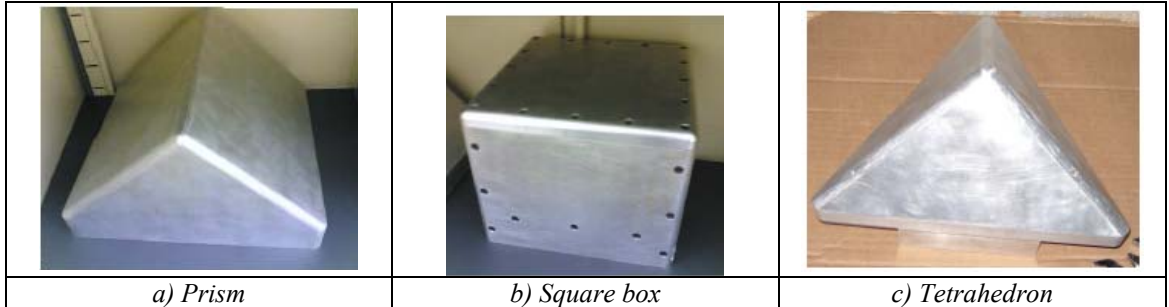

| *a) Prism* | *b) Square box* | *c) Tetrahedron* |

figure 1: punches

The second part of the device is constituted by a blank holder system: two mechanisms have been designed. The first one is a classical multi-part blank holder. It is composed of maximum 9 independent blank holders actuated by maximum 9 pneumatic jacks that enable to impose and sensor independently a variable pressure on each of them. Here again the system has been designed so that the contact zones of the black holders can be easily manufactured and changed to investigate the influence of their geometry on the process. The second disposal is designed to replace the blank holders by a fixed tension on the yarns extremities. This disposal is dedicated to investigate the real impact of the pressure and friction forces applied by the blank holders. It enables an easier comparison with finite element simulation since the complex phenomenon of pressure and friction between the yarns and the blank holders does not need to be modelled in this case.

The third part consists in a 3D Digital Image Correlation system. Two numerical cameras are located at the top of device, they can be positioned in function of the specific zone of the fabric that has to be analysed. The 3D displacement field is computed using a marker tracking software for markers plotted on the fabric sample at the initial state (figure 2). Much care has to be taken to plot the markers in order to be able to follow the yarns evolution.
With this 3D displacement field the membranous strains can be easily computed. The thickness variation is not so easy to extract from the displacement field and is not available at the moment, but it is under study.

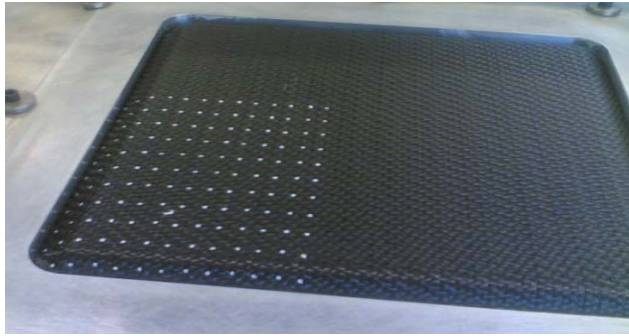
figure 2 : Sample with markers at the initial state

## MECHANICAL BEHAVIOUR OF THE FABRIC DURING FORMING

### Fabric used

The tests presented in this paper are carried out on a composite woven reinforcement used in aeronautics. It is denoted G1151® and constituted (figure 3) by an interlock weaving of 6K carbon yarns (630g/m², 7.5 yarns/cm).

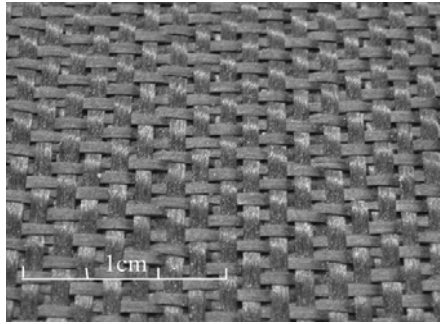
figure 3 : Woven carbon reinforcement (G1151®).

### Relationship between formability and mechanical behaviour

The main strain mode that enables fabrics to form on double curved parts is in plane shear. Thus identifying the shear behaviour of the fabric is crucial to be able to understand the forming behaviour of the fabric. The shear behaviour of the G1151 has been identified experimentally in our lab using both a specific picture frame and a bias-test. Details on these experiments and about the difficult problematic of realizing shear tests can be found in [6]. The shear curve obtained expresses the torque per area unit in function of the shear angle [6] and is presented on
figure 4.
The shear curve obtained is a classical one as far as in pane shear of fabrics is concerned [7]. The shear mechanism of fabrics is now quite well known and has been detailed using meso-optical analysis in [8]. The shear torque is very low until the lateral contact between the yarns of the same network starts to occur. Then the shear torque increases widely and round a specific angle that is called the "locking angle", the shear stiffness becomes too high and the fabric starts wrinkling. The shear angle should then not reach the locking angle value during forming in order to avoid wrinkling.

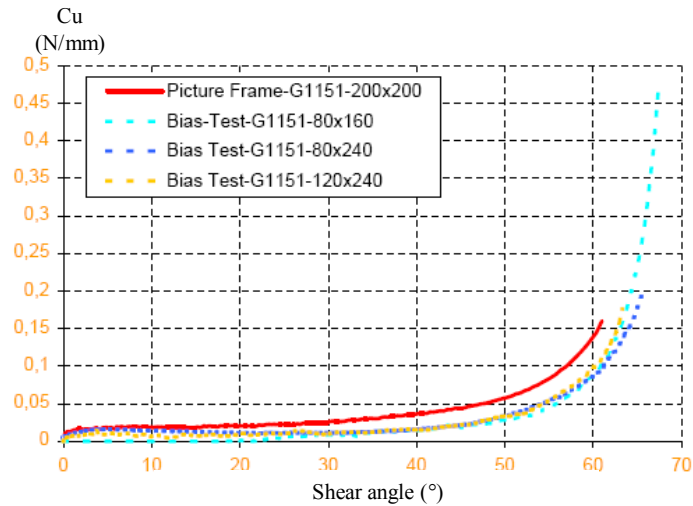

figure 4 : G1151 shear curve obtained using Picture Frame and Bias Test.

Obviously, the wrinkling phenomena: position, shape, number…, is strongly related to the bending stiffness of the layer. The bending behavour of the fabric has thus to be investigated. A specific device based on the generalization of the Pierce test and an optical measurements system has been developed in the Prisme Institute in order to characterize the bending behaviour of dry fabrics ([9], figure 5). Even if many works remain to do concerning our the understanding of the bending behaviour of fabrics, a hysteretic elastic plastic law expressing the bending torque in function of the curvature seems to be a good approach of the real bending behaviour measured. One curve used to identify the model is presented figure 5

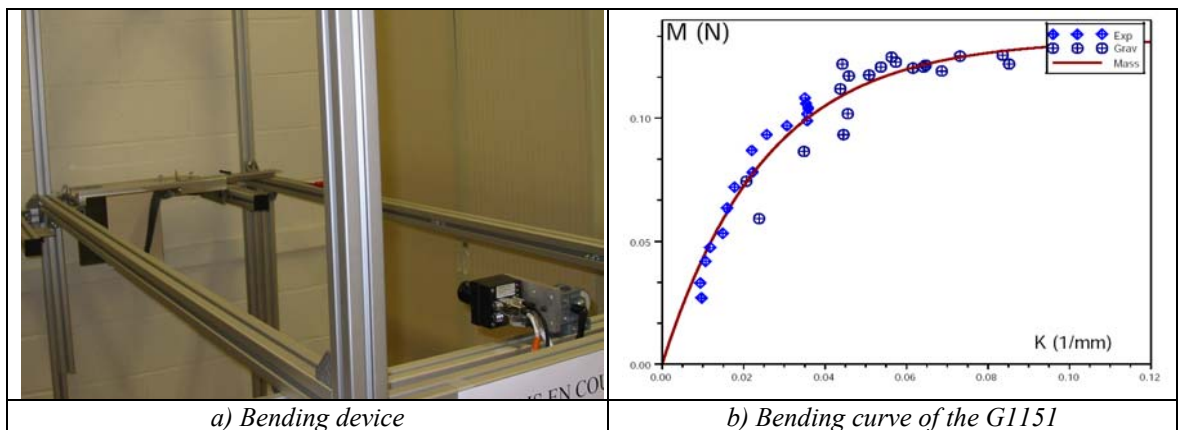

| *a) Bending device* | *b) Bending curve of the G1151* |

figure 5 : Bending device and bending curve of a layer of G1151

Another way to avoid wrinkling is to introduce tension in the yarns. Even if, to our knowledge, no quantitative influence of yarn tension on the locking angle is available, an increase of the locking angle has been observed by different teams during picture frame tests with pretension [6, 10]. Increasing the blank holder pressure enables to introduce tension in the yarns due to the friction between tools and fabric [11]. The blank holder should thus contribute to avoid wrinkling. But, since the tension will widely increase in the yarns, the biaxial behaviour of the fabric has to be identified in order to verify that the yarns are not damaged. The biaxial tensile surfaces [12] of the

G1151 have been measured on a specific biaxial tensile device developed in the lab [13] (figure 6). For more readability tensile surfaces can be expressed using two curves networks giving the load in one direction in function of the strain in the same direction and the strain ratio (figure 7).

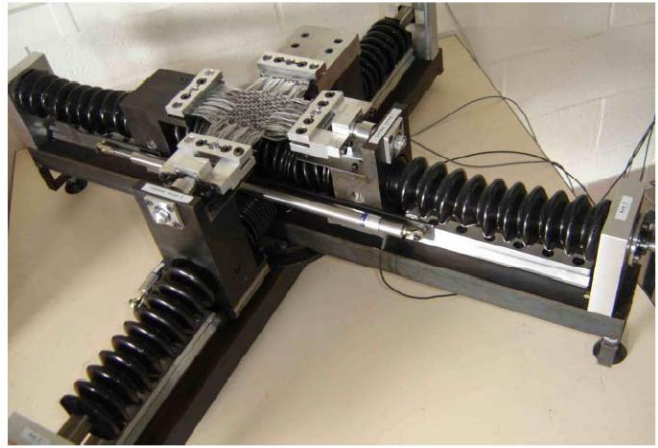

figure 6 biaxial tensile device

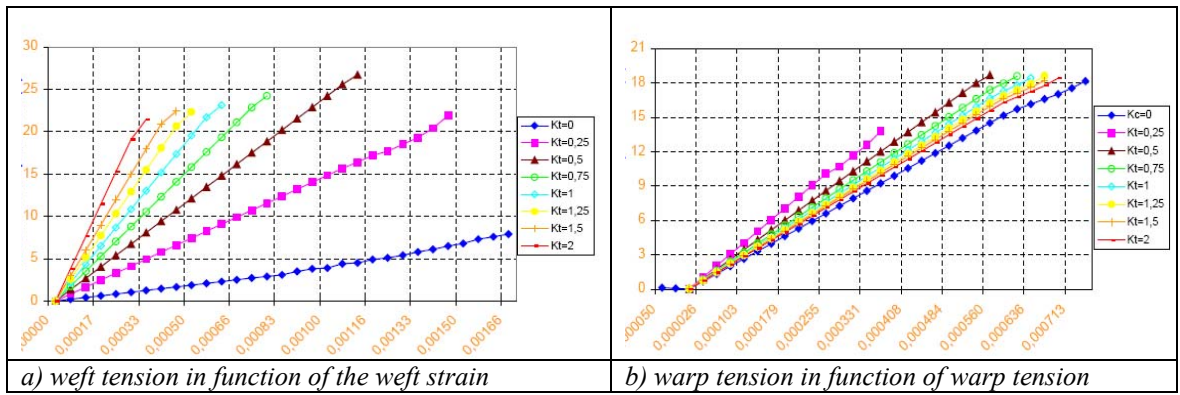

| *a) weft tension in function of the weft strain* | *b) warp tension in function of warp tension* |

figure 7 : Warp and weft tension in function of the biaxial strain ratio $Kt = \dfrac{\varepsilon_{warp}}{\varepsilon_{weft}}$

The displacement measured by the DIC system enables to obtain the membranous strains (shear angle and biaxial strains) and then to analyse them with respect to the mechanical behaviour measured.

**APPLICATION TO THE FORMING OF A SQUARE BOX**

figure 8 present the forming of a G1151 layer (1m²) using a square box punch (280x280x280 mm), edges radius value is 10mm. Due to this low value of the edges radius , the shape is highly double curved and its obtaining will need large strains of the fabric along the edges. A 1 bar pressure have been put on the 8 blank holders.
The global result of the stamping is presented figure 8. Yarns are oriented in the direction of the box at the initial state, so that they will be so at the deformed state.
In good agreement with the punch shape, the preform is symmetrical. We can observe that the reinforcement is well-stretched; on faces, edges or corner of this shape (figure 8).

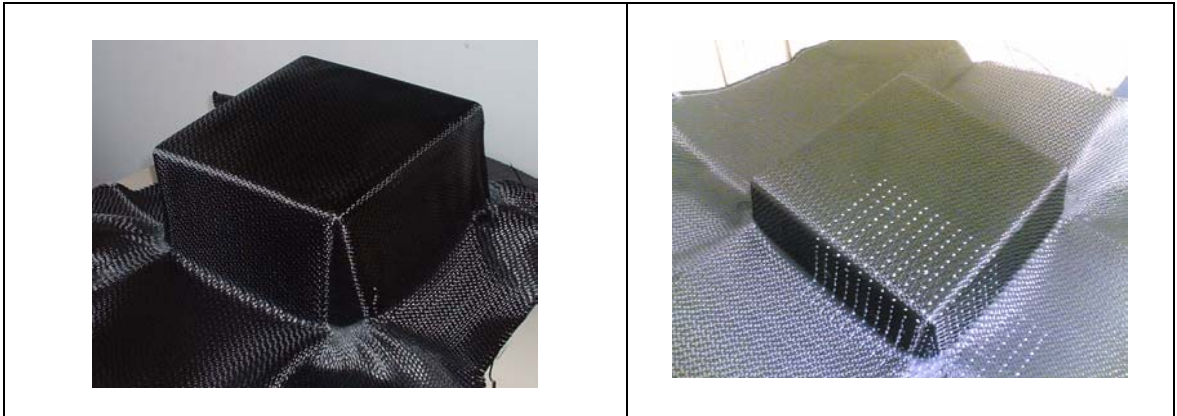

figure 8 : external view of the dry G1151 after the forming on the square box

As it has been explained in the previous paragraph, the instantaneous position of each marker plotted on fabric is obtained. The post processor reconstructs the 3D geometry of the preform in the different interesting zones (figure 9). As it was expected high shear angles can be observed on the edge of the box (50-60°, figure 9). However, no wrinkles can be observed in the useful zone of the final box. The process parameters (blank holders pressure, punch speed, die dimensions…) seem then to be suitable for the forming the given square box with the G1151.

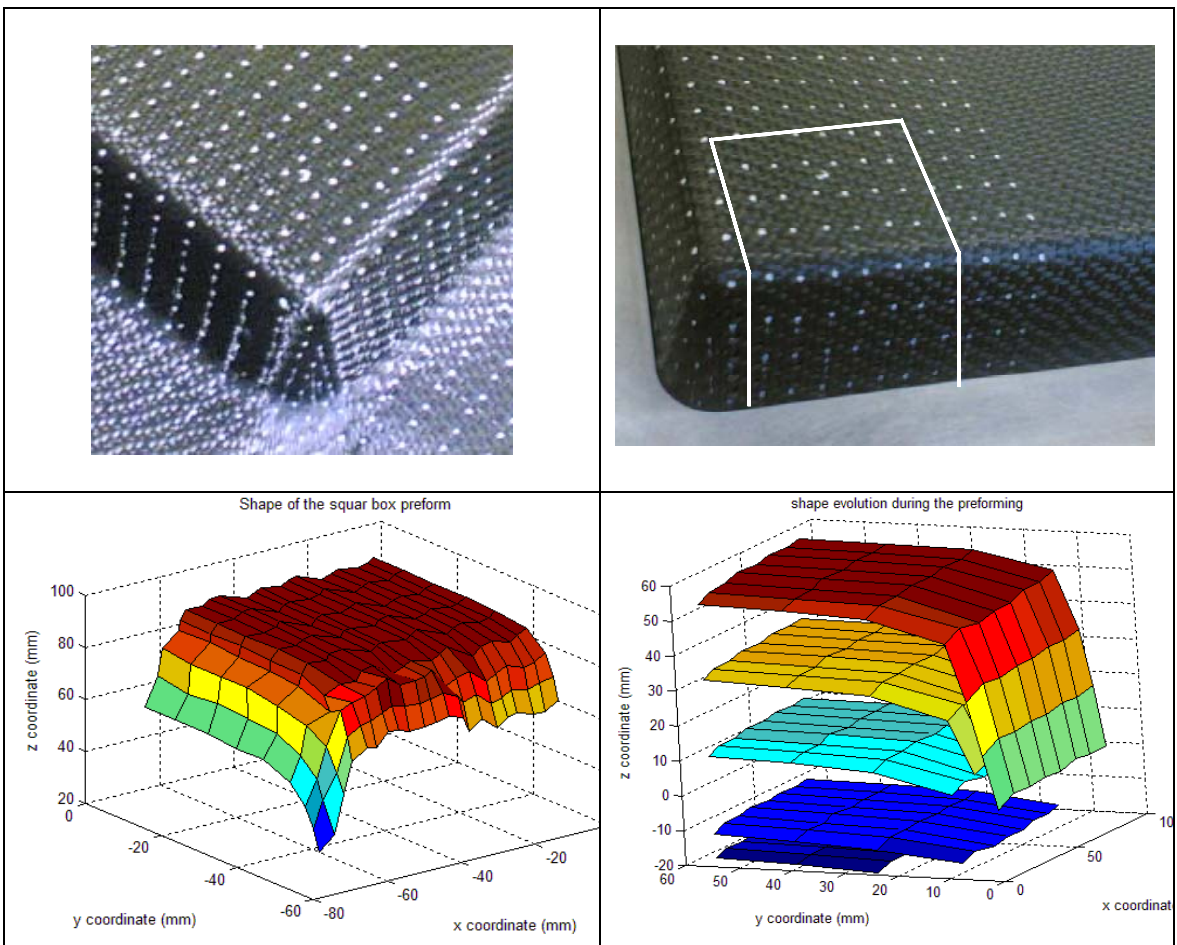

figure 9 : reconstruction of the deformed shape

# DEFECTS AFTER THE FORMING OF A THETRAEDRON PART

The second test that is presented in this paper is the forming of one layer with a tetrahedron punch. The process parameters are the following: 6 rectangular blank holders (area equal to 0.01m²), 1 bar pressure on each of them, 30mm/min punch speed, 160mm punch stroke. Different initial orientations of the fabric have been tested but only one will be presented here: the warp direction aligned with a symmetry plane of the tetrahedron. Results are presented on figure 10. Here again an outside view of the perform shows that the agreement with the punch shape at the end of the forming is pretty good. Wrinkles appear as it was foreseen but not in the useful part of the perform (figure 10 a). Nevertheless a closer look on the perform enables to see another type of defect that was not predicted by the analysis of the standard membranous tests on fabrics. In four specific zones (figure 10 b, figure 10c), buckles can be observed on the transverse yarns. These 4 zones are located in small bands that follow the direction of the longitudinal yarns and pass through the tetrahedron vertex (figure 10 b). It can be clearly seen (figure 10c) that transverse yarns are submitted to an out of plane bending. Their contribution of the biaxial stiffness of the membrane is then near 0, what is a real drawback for the final composites stiffness. These buckles are different from wrinkling since the membrane itself (the layer) is not wrinkled. Nevertheless, they are at least as problematic as wrinkles.

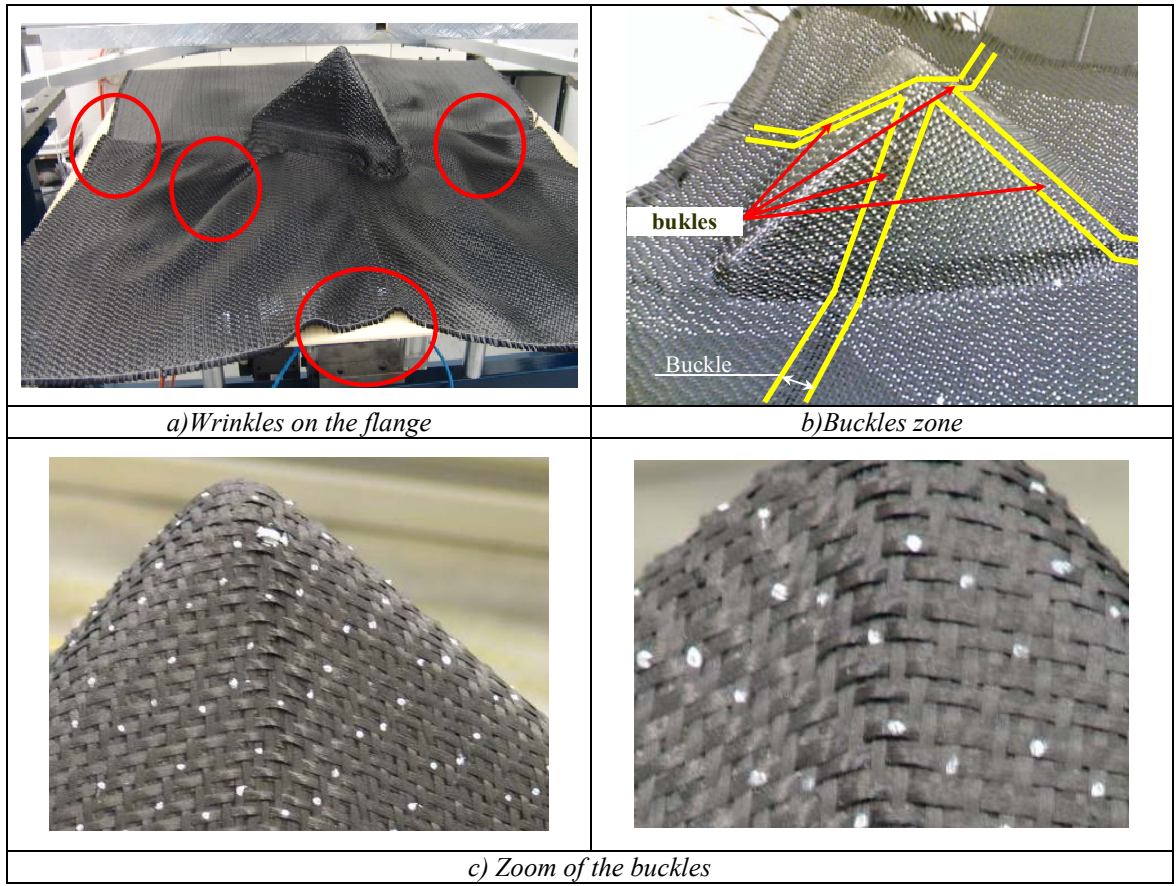

*a)Wrinkles on the flange* | *b)Buckles zone*

*c) Zoom of the buckles*

figure 10 : defects at the end of the forming of a layer of G1151 with a tetrahedron punch

**CONCLUSION**

The specific device designed through the collaboration of EADS and the PRISME Institute presented in this paper enabled to obtain first promising results. Instrumented with digital cameras, the displacement field can be reconstructed in order to analyze the link between the defects on the perform, the yarns orientation and the mechanical behaviour of the fabric that can be identified using specific experimental disposals designed in the PRISME Institute. The large panel of process conditions can be studied in order to see their influence on the perform obtained. This tool will undoubtedly permit to increase our knowledge on the forming process of dry fabrics. A type of defect (buckles) that was not attempted has, for example, already been pointed out for the tetrahedron shape. Many works have to be done in order to achieve our goal to define a priori the optimal process parameters and fabric properties. Furthermore, besides the experimental analysis of the process, this device will also enable to provide interesting data in order to develop the numerical models.

**ACKNOWLEDGEMENTS**

An amicable thought and many thanks to Fabric Labbe who has strongly contributed to the design of the device.